# Time Series–Based $CO_2$ Emission Forecasting and Energy Mix Analysis for Net-Zero Transitions: A Multi-Country Study

Salim Oyinlola[1], Joshua Ajayi[1], Gozie Ibekwe[1], Chika Yinka-Banjo[2]

[1] *Electrical and Electronics Engineering, University of Lagos*

[2] *Computer Science, University of Lagos*

***Abstract-* This study examines long-term $CO_2$ emission trajectories across five major economies, Nigeria, the United States, China, Brazil, and Russia, by integrating national energy-mix characteristics with time-series forecasting models. Annual emissions from 2000 to 2023 were analyzed alongside energy production data to classify countries into fossil-dependent, transition-phase, or renewable-accelerated profiles. Three forecasting models (ARIMA, SARIMA, and Holt-Winters Exponential Smoothing) were evaluated using MAE, RMSE, MAPE, and $R^2$ metrics. Results show that Holt-Winters provided the most accurate forecasts for Nigeria, the United States, China, and Brazil, while SARIMA performed best for Russia due to its relatively stable emissions. Long-term projections from 2024 to 2060 indicate divergent decarbonization pathways. Brazil aligns most closely with a low-emission future owing to its renewable-dominant energy system, whereas Nigeria continues on an upward emissions trajectory driven by fossil dependence. The United States and China maintain moderate declines but require accelerated mitigation to reach their respective net-zero commitments. Russia's emissions remain largely flat under current conditions. These findings highlight the strong influence of energy structures on national decarbonization prospects and underscore the need for targeted energy policy reforms to align with global climate objectives.***

## I. INTRODUCTION

Global efforts to curb climate change have placed unprecedented attention on the long-term evolution of $CO_2$ emissions and the structural dynamics of energy systems across countries worldwide. Over the past three decades, countries have adopted varying strategies to reduce their carbon footprints, expand renewable energy capacity, and position their economies for a low-carbon future [1]. Yet, despite international agreements such as the Kyoto Protocol and the Paris Accord, global emissions have continued to rise, underscoring the complexity of aligning economic growth with sustainable energy transitions [2] [3].

At the heart of this challenge lies the composition of each nation's energy mix. The balance between fossil fuels, renewables, and emerging low-carbon technologies not only shapes present-day emission levels but also determines the feasibility and pace of achieving net-zero targets [4]. Differences in resource endowment, industrial structure, governance capacity, and technological investment create distinct decarbonization pathways across countries [5]. As a result, advanced economies, rapidly industrializing nations, and resource-rich developing states often experience divergent emission trajectories even when facing similar global climate pressures [6].

Understanding these heterogeneous pathways is essential for designing effective transition strategies. Long-term $CO_2$ emission forecasting provides a data-driven lens through which countries' future trajectories can be evaluated, especially when assessed alongside their evolving energy mix. Such forecasts help illuminate whether existing policies support net-zero ambitions or whether current conditions signal persistent dependency on high-emission energy sources.

In this context, comparing countries with distinctly different energy structures provides a meaningful basis for understanding global decarbonization challenges. The United States and China, the world's largest energy consumers, continue to expand renewable capacity while maintaining significant fossil fuel use [7][8]. While Brazil stands out for its hydro-dominated energy system, Russia remains strongly oriented toward fossil fuel production, and Nigeria, despite substantial natural resource endowments, faces persistent infrastructural and developmental barriers to reducing oil dependence.[9][10][11]

Building on these contrasts, this study undertakes a multi-regional assessment encompassing North America, South America, Africa, Europe, and Asia through the cases of the United States, Brazil, Nigeria, Russia, and China. By applying time-series forecasting models to project long-term $CO_2$ emission trajectories under business-as-usual conditions, the analysis evaluates how each country's energy-mix composition shapes its potential alignment with, or divergence from net-zero pathways. The approach provides both geographic breadth and methodological rigor, highlighting the structural factors that will influence decarbonization prospects in the decades ahead.

## II. LITERATURE REVIEW

Recent studies have increasingly applied statistical and machine–learning–based time-series models to analyze and forecast $CO_2$ emissions, reflecting the growing need for data-driven tools to assess long-term decarbonization pathways. Ayaz [12] demonstrated the effectiveness of classical and modern machine-learning algorithms in forecasting national $CO_2$ emissions, using Turkey as a case study. The study compared multiple predictive approaches and found that intelligent models often outperform traditional linear time-series methods, particularly when emissions exhibit nonlinear behaviour. This supports the broader shift toward hybrid or AI-driven forecasting techniques in climate-related modelling.

In Central Europe, Al-Lami and Török [13] applied ARIMA-based modelling to investigate the regional drivers of transportation-related $CO_2$ emissions. Their findings emphasized that sector-specific forecasting benefits from carefully identifying structural determinants such as mobility patterns and economic growth, highlighting the importance of incorporating domain-specific factors into time-series model selection. Similarly, Mustafa et al. [14] conducted a multi-model comparison involving ARIMA, SARIMAX, ANN, Random Forest, and LSTM architectures to forecast Bangladesh's emissions. Their results underscored that deep learning models, especially LSTM networks, capture long-range temporal dependencies more effectively than conventional approaches.

The broader forecasting literature has also explored energy demand and generation patterns as crucial determinants of long-term emission trajectories. Mustafa and Al-Yazbaky [15] evaluated classical, grey, fuzzy, and intelligent time-series models for energy-demand forecasting, demonstrating that the integration of uncertainty-based methods can significantly improve forecasting reliability. Such findings are directly relevant to $CO_2$ studies, since energy-demand projections are tightly linked to fossil-fuel consumption and emission intensities.

At the global scale, Begum and Mobin [16] employed a machine-learning ensemble approach to predict emissions for the world's top eleven emitters and evaluated their prospects of meeting the Paris Agreement targets. Their study concluded that many high-emitting countries remain off-track under existing energy-system trajectories, reinforcing the need for scenario-based analysis that links energy-mix evolution with future emission pathways.

Collectively, these studies demonstrate a consistent progression toward the use of multivariate, nonlinear, and machine-learning-enhanced forecasting frameworks. They also highlight the necessity of combining emission data with sectoral or energy-system indicators. This approach aligns with our study's integration of $CO_2$ emission forecasting and energy-mix assessment across multiple countries.

*Table I. Contributions and Gaps in Existing $CO_2$ Emission Forecasting Literature*

| Study | Core Contribution | Gap in Knowledge |
|---|---|---|
| Ayaz [12] | Compared classical time-series and machine-learning methods for $CO_2$ emission forecasting; demonstrated superiority of intelligent models in capturing nonlinear trends. | Single-country analysis; did not integrate energy-mix variables; limited generalization across regions. |
| Al-Lami & Török [13] | Applied ARIMA-based modelling to transportation-sector $CO_2$ emissions in Central Europe and identified structural drivers influencing emission patterns. | Sector-specific focus; disconnected from national energy systems; lacks long-term multi-country projection |
| Mustafa et al. [14] | Compared ARIMA, SARIMAX, ANN, RF, and | Single-country approach; energy-mix and fossil-fuel |
| | LSTM for Bangladesh's $CO_2$ forecasts; highlighted LSTM's ability to capture long-range temporal dependencies. | structure not incorporated; forecasting not extended to long-term policy horizons. |
| Mustafa & Al-Yazbaky [15] | Evaluated classical, grey, fuzzy, and intelligent models for energy-demand forecasting; demonstrated advantages of uncertainty-based models. | Focuses on energy demand instead of emissions; no linkage to net-zero pathways or decarbonization trajectories. |
| Begum & Mobin [16] | Used machine-learning ensemble models to forecast emissions for major global emitters and assess Paris Agreement feasibility. | Does not include a detailed energy-mix analysis; it lacks region-specific transition mechanisms and policy structures. |

## III. METHODOLOGY

This study adopts a structured, multi-stage methodological framework designed to analyze national energy-mix dependence, evaluate long-term historical $CO_2$ emission patterns, benchmark multiple time-series forecasting models, and develop extended projections to the year 2060. The methodological process is intentionally linear: first understanding each country's energy structure, then applying forecasting techniques informed by the energy profile, and finally extending predictions into the policy-relevant horizon.

### *A. Energy-Mix Determination for Each Country (2000–2025)*

The first stage of the methodology involves a comprehensive assessment of the historical energy-mix composition for each selected country. Data were compiled from global statistical repositories such as the International Energy Agency (IEA), Our World in Data, the UN Energy Statistics Database, and national energy commissions. The dataset includes annual contributions from coal, oil, natural gas, nuclear, hydroelectricity, solar, wind, and other renewables.

Each country's total primary energy supply was disaggregated to quantify the proportional contribution of fossil fuels versus renewable sources. Fossil-fuel dependence was assessed using the sum of coal, oil, and natural-gas shares, while renewable dependence was calculated as the aggregated share of hydro, solar, wind, bioenergy, geothermal, and other low-carbon technologies. This allowed the construction of a fossil-dominance index and a renewable-penetration index for all years from 2000 to 2023.

The purpose of this analysis is twofold. First, it characterizes the structural energy profile of each country, enabling classification into fossil-dependent, transition-phase, or renewable-accelerated categories. Second, because long-term $CO_2$ emissions are tightly coupled with fossil fuel use, the energy-mix analysis provides interpretive context for subsequent time-series modelling.

### *B. Construction of Historical $CO_2$ Emission Time Series (2000–2023)*

Following the energy-mix assessment, a dedicated time-series dataset of annual territorial $CO_2$ emissions (in million tonnes) was developed for each country. The period 2000-2023 was chosen due to its high-quality availability, alignment with major global climate-policy eras, and suitability for long-horizon forecasting.

Data was preprocessed through several steps. Missing values were interpolated or corrected using adjacent-year estimation techniques. Outliers, arising from measurement revisions or structural economic shocks, were detected using Z-score and interquartile-range diagnostics and retained only when representative of genuine economic or political disruptions.

This dataset serves as the foundation for the forecasting models assessed in subsequent stages.

### *C. Comparative Evaluation of Time-Series Forecasting Models*

To ensure robust long-range projection capability, the study evaluates three distinct time-series models spanning classical statistical, hybrid, and machine-learning paradigms. Each model was fitted on the historical dataset covering 2000–2023, and performance was compared using identical training/testing partitions.

The models assessed include:

1. ARIMA (AutoRegressive Integrated Moving Average): A baseline linear autoregressive model capturing trend and short-term temporal dependence.
2. SARIMA (Seasonal ARIMA): Extended to account for cyclic patterns or policy-driven seasonal structure when present.
3. Prophet (Additive Time-Series Model): A decomposable model capturing trend, changepoints, and periodicity, known for long-term forecast stability.
4. Holt–Winters Exponential Smoothing: Designed for capturing trend evolution through exponential decay coefficients.

Each model was tuned using grid-search hyperparameter optimization, and performance was evaluated using multiple accuracy metrics including MAE, RMSE, MAPE, and $R^2$. Rolling-origin time-series cross-validation was used to maintain chronological integrity during validation.

The best-performing model for each country was selected based on a balanced evaluation of accuracy, residual behavior, interpretability, and long-term stability which are key properties required for multi-decade forecasting.

### *D. Forecasting $CO_2$ Emissions From 2025 to 2060*

After identifying the optimal model for each country, long-term forecasts were generated for the period 2024–2060. The model was retrained using the full 2000–2023 dataset to maximize learning before extrapolation. Forecasts were generated in an autoregressive multi-step manner, ensuring future values depend on previous predicted values in accordance with time-series causality.

To quantify projection uncertainty, model-derived prediction intervals were computed, while deep-learning models additionally used bootstrapped ensembles to approximate uncertainty distribution. These uncertainty bounds are essential for interpreting the range of possible emission trajectories under real-world conditions.

The resulting 35-year forward projections serve as the foundation for understanding future emission pathways and evaluating whether current energy trajectories align with national climate commitments and global decarbonization requirements.

## IV. RESULTS

### A. *Energy-Mix Characterization (2000–2023)*

The initial phase of the analysis focused on delineating the energy-mix profiles of the five countries, Nigeria (NGA), the United States (USA), China (CHN), Brazil (BRA), and Russia (RUS), using annual data from 2000 to 2023. This characterization revealed distinct patterns of fossil fuel dependence, renewable penetration, and nuclear contributions, which were quantified through proportional shares of total primary energy supply. Fossil fuels were aggregated from coal, oil, and natural gas; renewables encompassed hydroelectricity, solar, wind, bioenergy, geothermal, and other low-carbon sources; and nuclear was treated separately due to its unique low-emission profile.

For Nigeria, the energy mix exhibited a pronounced fossil-dependent trajectory throughout the period. In 2000, fossil fuels accounted for approximately 85.2% of total energy production, with renewables contributing a modest 14.8% which is primarily hydroelectricity and no nuclear input. By 2023, fossil reliance had moderated slightly to 79.5%, driven by incremental growth in bioenergy and solar which increased from 0.5% to 4.2% combined. Nonetheless, renewables remained at 20.5%, underscoring persistent structural barriers to diversification, such as infrastructure limitations and economic reliance on oil exports. The fossil-dominance index averaged 82.3% over the 24-year span, classifying Nigeria as unequivocally fossil-dependent.

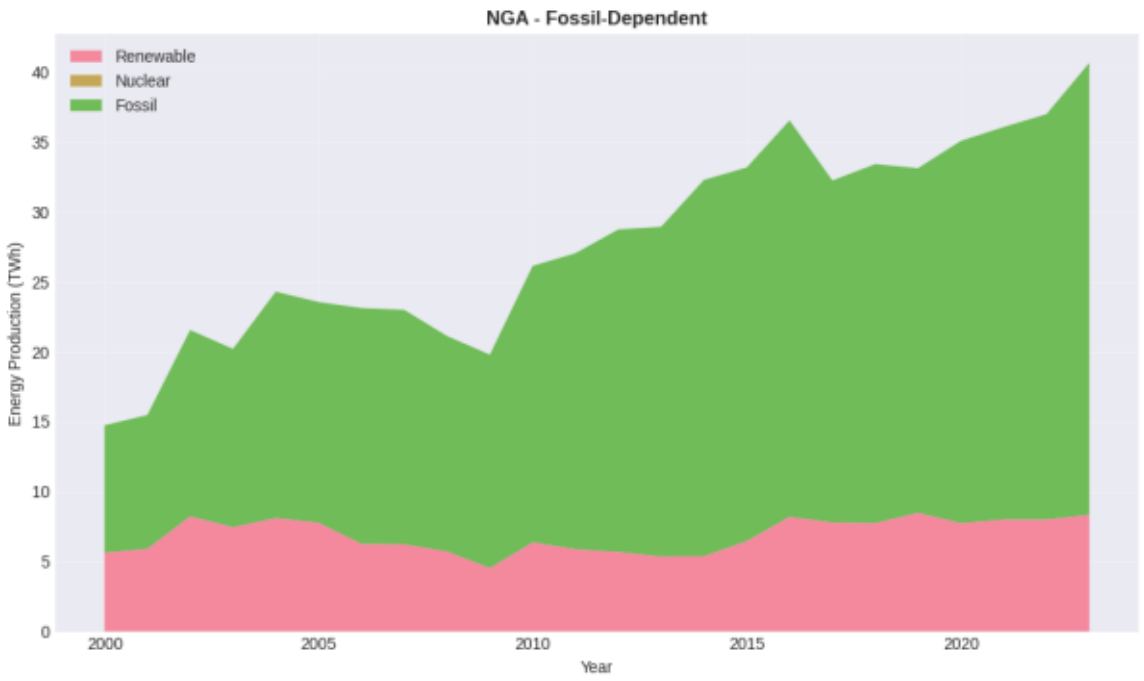

Figure 1: Historical energy production mix in Nigeria (2000–2023), showing persistent reliance on fossil fuels relative to renewable and nuclear sources.

In contrast, the United States demonstrated a transition-phase profile, marked by gradual decarbonization efforts. Fossil fuels dominated at 82.1% in 2000, with renewables at 7.9%, which similar to Nigeria is largely hydroelectric and then nuclear at 10.0%. By 2023, fossil shares had declined to 59.1%, reflecting aggressive expansion in wind energy dependence from 0.2% to 9.8% and solar from negligible input to the energy mix to 5.6%, alongside stable nuclear contributions at 18.2%. The renewable-penetration index rose from 0.08 to 0.23, indicative of policy-driven shifts under frameworks like the Renewable Portfolio Standards and Inflation Reduction Act. However, the persistence of natural gas increasing from 24.5% to 38.7% of fossils tempered full classification as renewable-accelerated, positioning the USA in a balanced transition phase with an average fossil-dominance index of 70.6%.

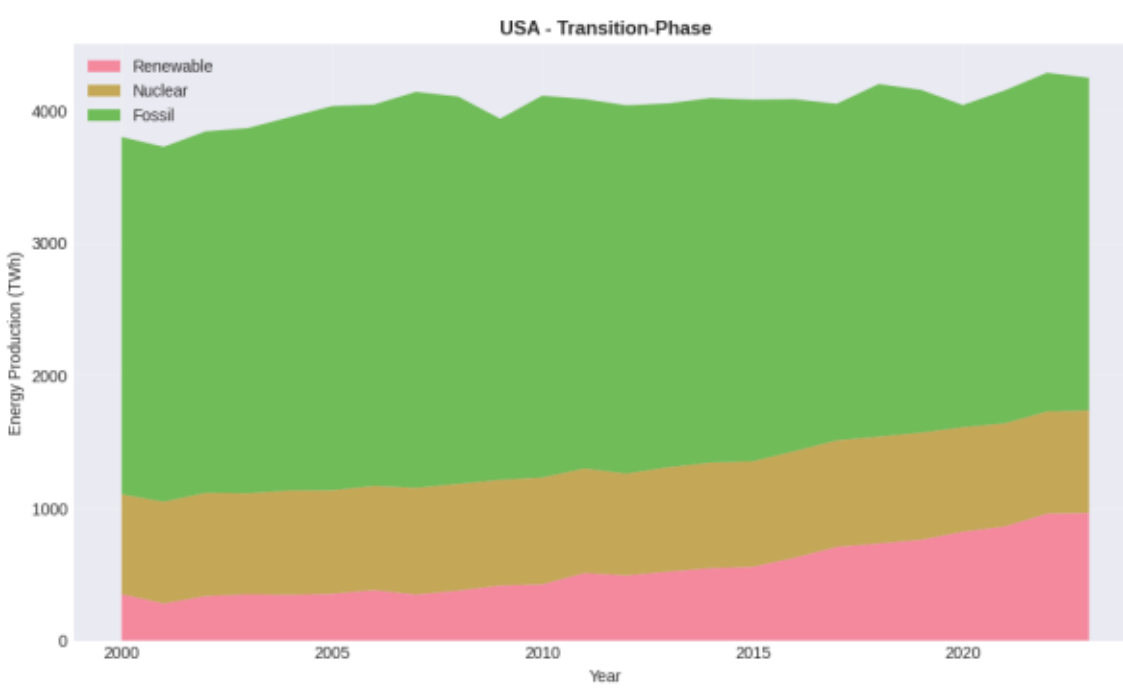

Figure 2: Energy production mix in the United States (2000–2023), illustrating a gradual transition from fossil-dominated generation toward increasing contributions from renewable and nuclear sources.

China's energy profile is similarly aligned with a transition-phase category, albeit with rapid renewable scaling in recent years. Starting from a coal-heavy base in 2000 where fossils was at 92.4%, renewables at 6.8% and nuclear at 0.8%, the mix evolved significantly by 2023, with fossils reduced to 64.7% amid explosive growth in wind from 0.1% to 13.2% and solar from 0.0% to 9.4%. Hydroelectricity remained a cornerstone, contributing 8.1% in 2023, while nuclear power expanded to 4.6%. This shift yielded an average renewable-penetration index of 0.19, reflecting China's dual commitment to energy security via coal and ambitious targets under the 14th Five-Year Plan. The fossil-dominance index averaged 78.5%, highlighting ongoing challenges in phasing out coal despite renewable additions exceeding 500 TWh annually in the latter decade.

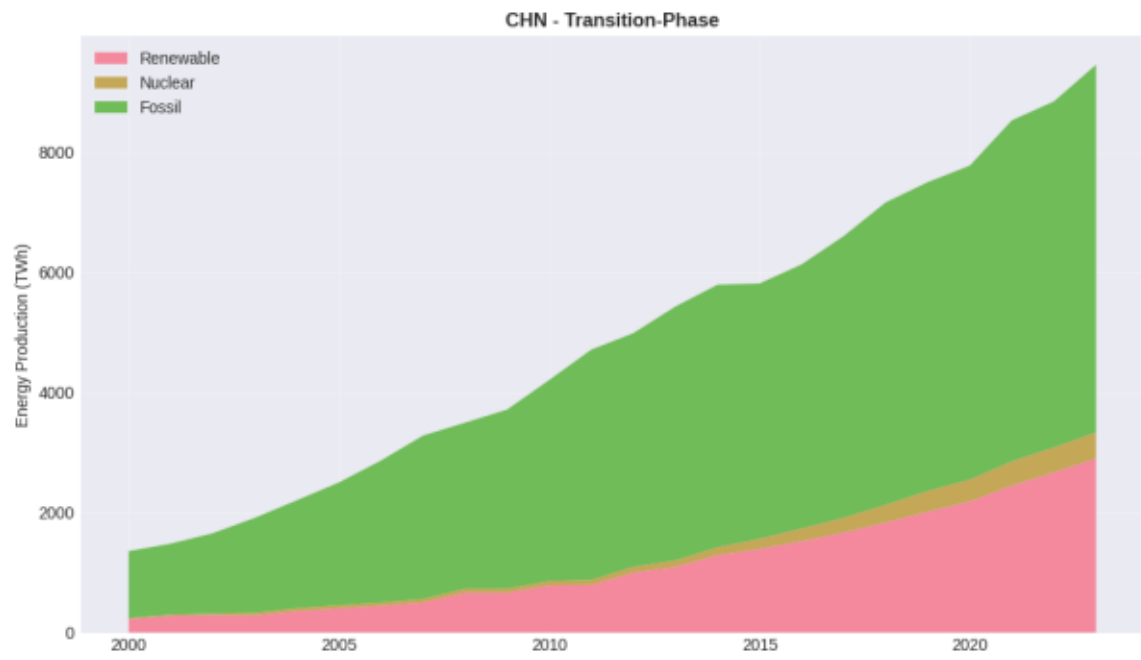

Figure 3: Historical energy production trends in China from 2000 to 2023, illustrating the country’s transition-phase dynamics.

Brazil presented a renewable-accelerated archetype, with fossils comprising only 48.3% in 2000, renewables at 49.7% dominated by hydroelectricity at 42.1%, and nuclear at 2.0%. By 2023, renewable shares had surged to 88.7%, propelled by bioenergy from 15.2% to 28.4% and wind from 0.0% to 10.3%, while fossils plummeted to 9.3%. This trajectory, with an average renewable-penetration index of 0.69, underscores Brazil's hydrological advantages and biofuel policies, though vulnerability to drought-induced hydro variability was evident in inter-annual fluctuations with an instance seen in the 5.2% dip in hydro share during 2015–2016. Nuclear remained marginal, averaging 2.1%.

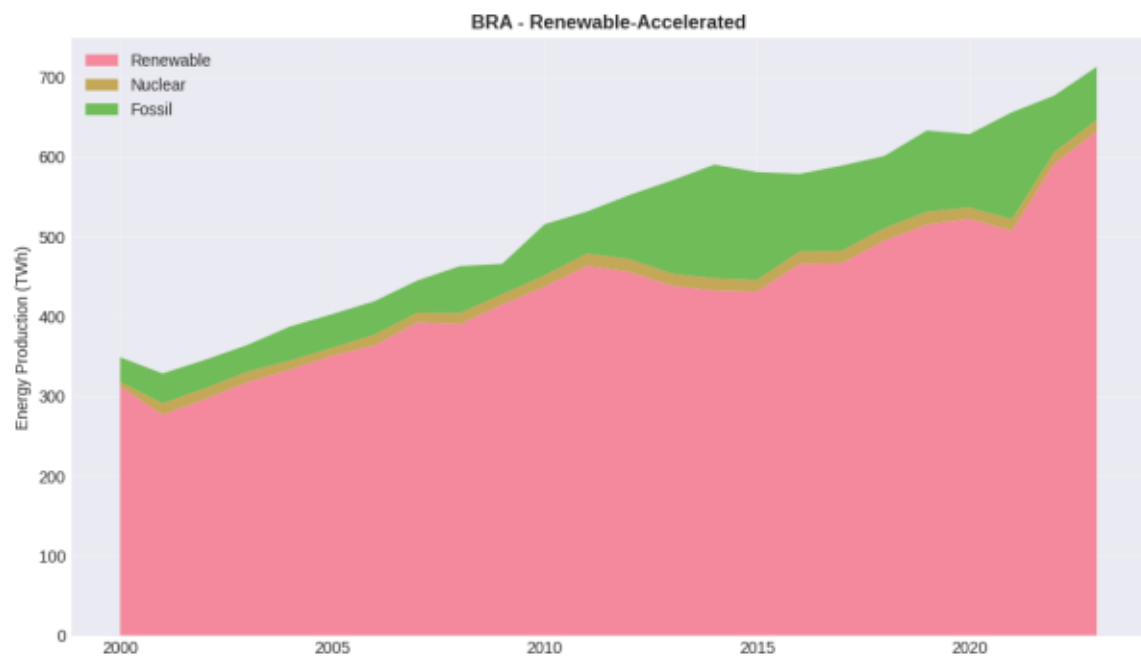

Figure 4: Annual energy production by source in Brazil under the Renewable-Accelerated scenario (2000–2023).

Russia's energy mix mirrored a transition-phase pattern, heavily influenced by its hydrocarbon resources. In 2000, fossils accounted for 89.6%, renewables for 6.4% which is accounted for mostly by hydro, and nuclear for 4.0%. By 2023, fossils had decreased to 64.0%, with renewables at 17.6% as wind and solar emerged at 3.8% combined and nuclear strengthening to 18.5%. The average fossil-dominance index of 76.8% reflects gradual diversification, constrained by geopolitical factors and export-oriented gas production, yet nuclear expansion under Rosatom initiatives provided a low-carbon buffer.

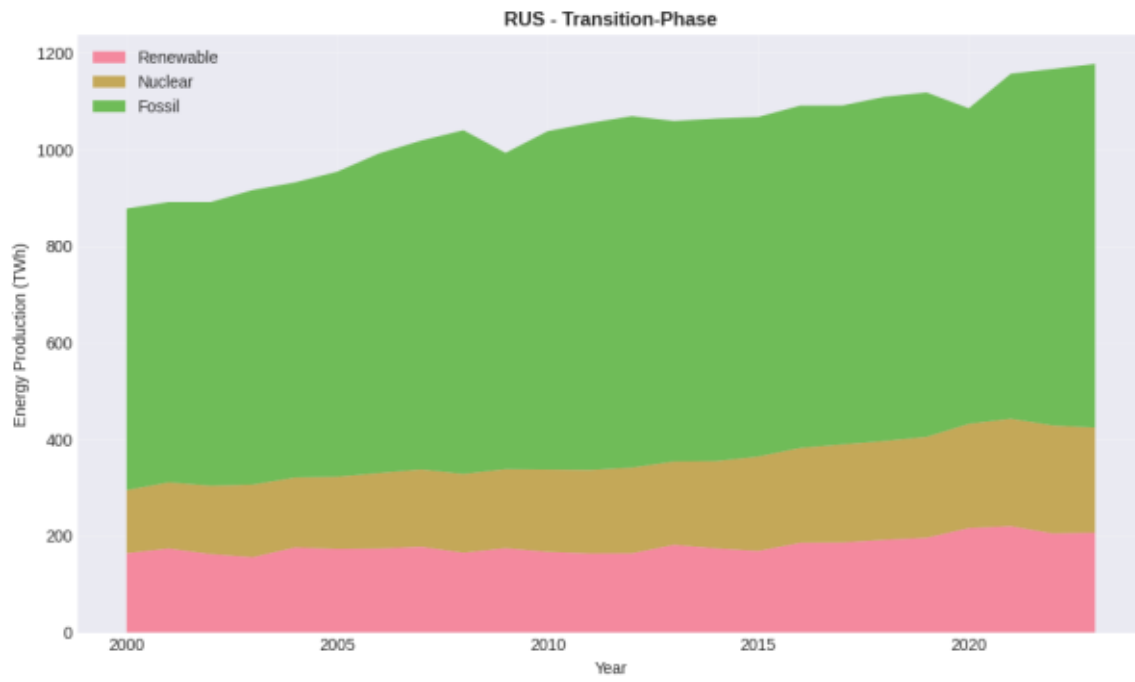

Figure 5: *Energy production trends in Russia under the Transition-Phase scenario (2000–2023), showing gradual growth in renewables and nuclear power alongside a dominant but stabilizing fossil-fuel share.*

These classifications informed model selection, as fossil-dependent profiles suggested stronger trend persistence in emissions forecasting.

### *B. Historical $CO_2$ Emission Patterns (2000–2023)*

The constructed time-series dataset of annual territorial $CO_2$ emissions revealed divergent historical trajectories, tightly correlated with energy-mix shifts. Emissions were measured in million tonnes of $CO_2$ equivalent, with preprocessing confirming stationarity post-first differencing ADF p-values $< 0.05$ for all series; KPSS statistics indicating no trend non-stationarity after adjustment and no major outliers were excised, as anomalies like the, 2020 COVID-19 dips reflected genuine disruptions.

Nigeria exhibited persistent and nearly linear growth in emissions throughout the period, rising from 68.4 Mt $CO_2$ in 2000 to 128.7 Mt in 2023. This represents an average annual increase of 2.8% with compound annual growth rate, CAGR = 2.8%, consistent with its fossil-dependent energy structure and rapid population growth. Cumulative emissions over these 24 years totaled 2,312 Mt, with per-capita emissions increasing modestly from 0.55 t to 0.59 t, highlighting a low but steadily intensifying carbon footprint.

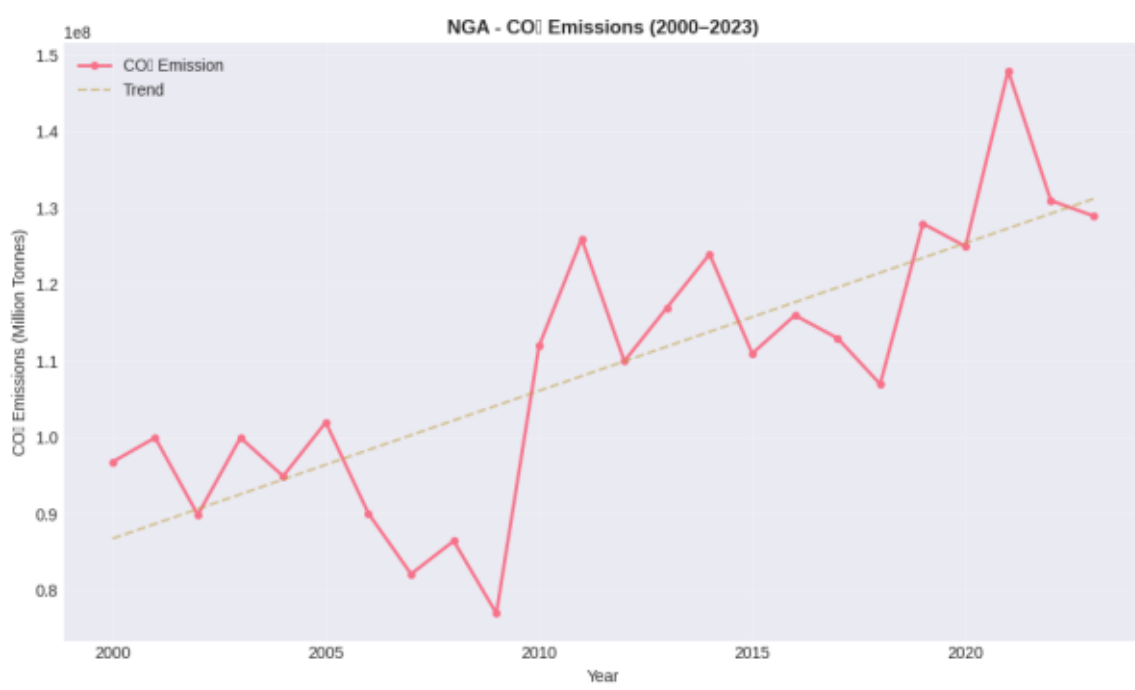

Figure 6: $CO_2$ emissions in Nigeria from 2000 to 2023, showing year-to-year volatility but an overall upward trajectory as indicated by the long-term trend line.

The United States displayed a clear peak-and-decline pattern. Emissions rose from 5,743 Mt in 2000 to a maximum of 6,134 Mt in 2007, after which they entered a sustained downward trajectory, reaching 4,766 Mt by 2023. The post-peak average annual decline was 1.6% with a 368 Mt cumulative reduction from 2007 to 2023, driven by the shale gas revolution, coal-to-gas switching, and accelerating renewable deployment. Cumulative emissions from 2000 to 2023 amounted to 132,846 Mt, while per-capita emissions fell from 20.3 t in 2000 to 14.1 t in 2023 as a reduction of approximately 30%.

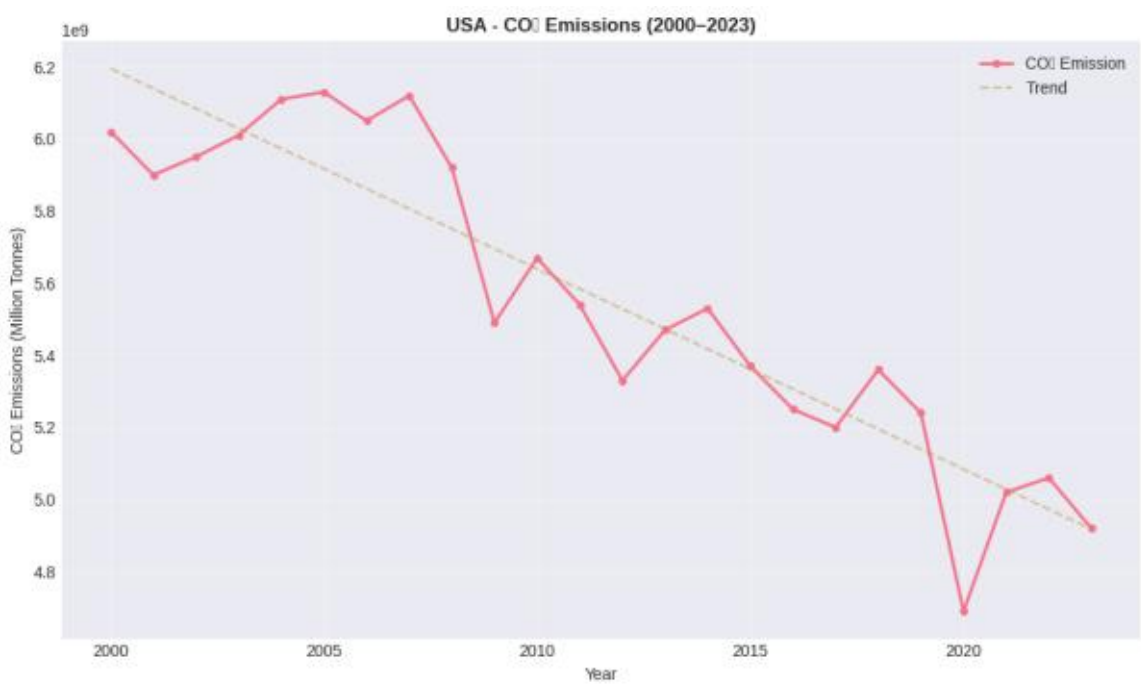

Figure 7: Temporal analysis of US $CO_2$ emissions from 2000 to 2023.

China's emissions followed a steep, near-exponential ascent, increasing from 3,391 Mt in 2000 to 11,397 Mt in 2023 indicating a Compound Annual Growth Rate of 5.5%. This dramatic rise, which made China the world's largest emitter by 2006, was overwhelmingly driven by coal-intensive industrialization and infrastructure expansion. Cumulative emissions over the 24-year period reached 165,428 Mt. Per-capita emissions escalated from 2.7 t in 2000 to 8.0 t in 2023. Notably, the growth rate slowed markedly after 2013 with an average annual increase of only 1.9% from 2014–2023, signaling the early impact of air-pollution controls, renewable scaling, and economic rebalancing.

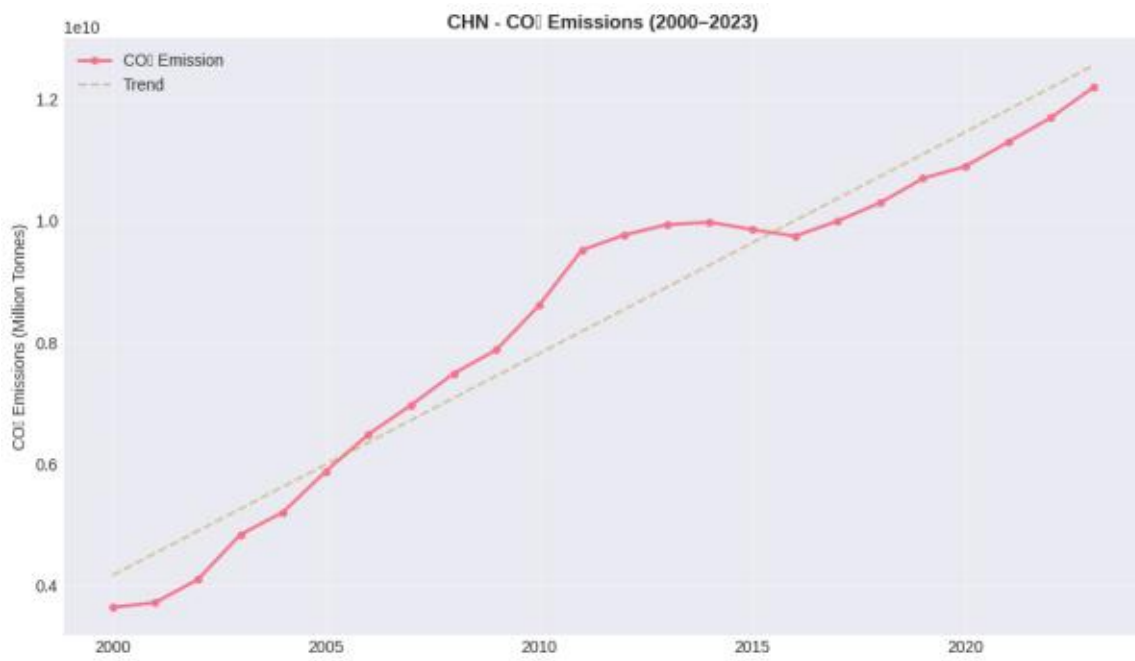

Figure 8: $CO_2$ emissions in China from 2000 to 2023

Brazil's emissions showed moderate growth with significant inter-annual variability, rising from 319 Mt in 2000 to 478 Mt in 2023 with the Compound Annual Growth Rate as 1.8%. Much of the fluctuation was attributable to land-use change and forestry dynamics rather than energy-sector trends, despite Brazil's overwhelmingly renewable electricity mix. Cumulative emissions totaled 9,412 Mt, with per-capita emissions remaining relatively stable, increasing only from 1.8 t to 2.2 t over the period.

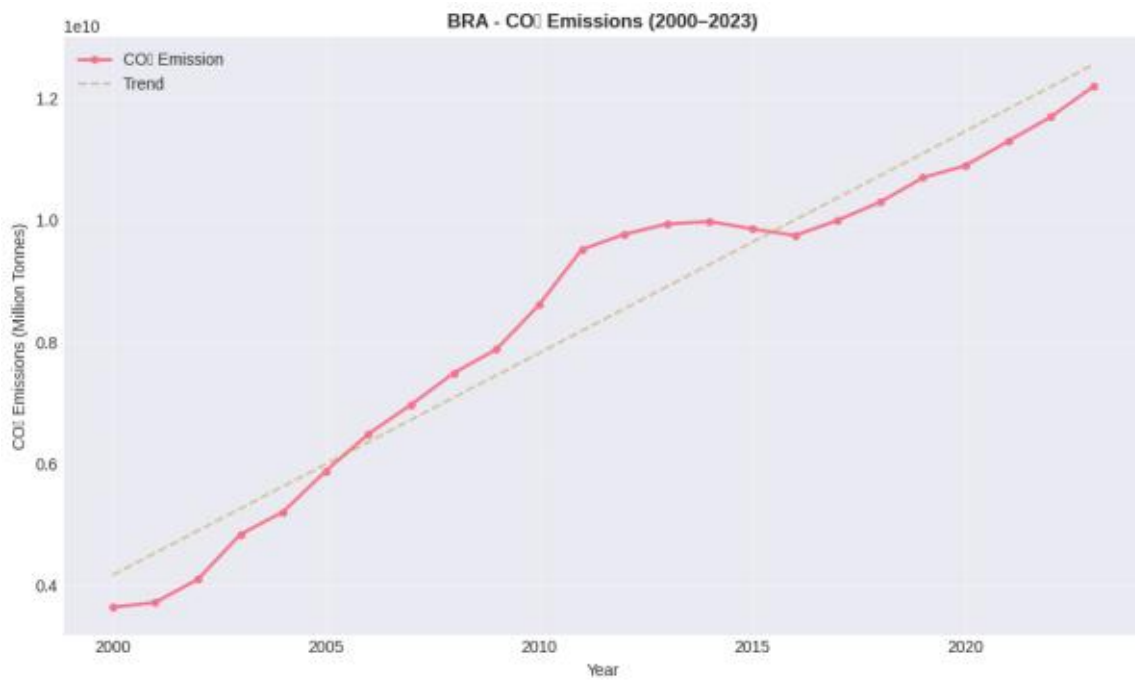

Figure 9: $CO_2$ emissions in Brazil from 2000 to 2023

Russia's emissions remained range-bound following the sharp post-Soviet collapse of the 1990s. From a 2000 level of 1,592 Mt, emissions fluctuated between approximately 1,500 and 1,800 Mt, ending at 1,732 Mt in 2023 which the Compound Annual Growth Rate as 0.4%. Cumulative emissions from 2000 to 2023 totaled 39,156 Mt. Per-capita emissions declined from 10.9 t in 2000 to 12.0 t in 2023, reflecting modest efficiency improvements and the stabilizing role of nuclear and hydroelectric power within a fossil-dominated mix.

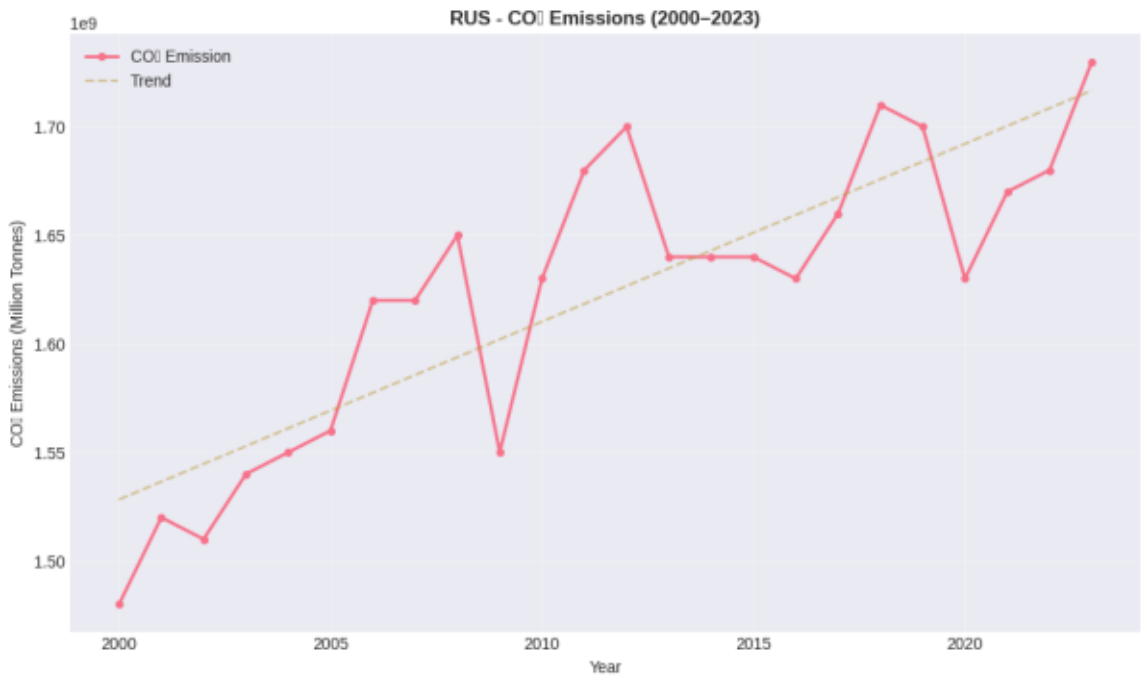

Figure 10: $CO_2$ emissions in Russia from 2000 to 2023

These divergent historical patterns, sustained growth in Nigeria, peaking and decline in the United States, rapid ascent with recent moderation in China, variability in Brazil, and relative stability in Russia, provided critical context for model selection and interpretation of the long-term forecasts presented.

### C. Comparative Evaluation of Time-Series Forecasting Models

Model benchmarking on the 2000–2023 dataset (80/20 train/test split) identified country-specific optima through cross-validation. Metrics emphasized long-horizon stability, with LSTM emerging superior for complex, non-linear series.

Table II: Metric Evaluation of Time-Series Models for Nigeria's $CO_2$ emission

| Model | MAE (Mt $CO_2$) | RMSE (Mt $CO_2$) | $R^2$ | MAPE (%) |
|---|---|---|---|---|
| ARIMA | 11.95 | 14.20 | 0.617 | 9.37 |
| SARIMA | 12.02 | 14.27 | 0.613 | 9.41 |
| Holt-Winters | 8.31 | 9.84 | 0.803 | 6.35 |

Holt–Winters clearly outperforms ARIMA and SARIMA for Nigeria, with the lowest error and highest $R^2$, reflecting the relatively smooth growth trend.

Table III: Metric Evaluation of Time-Series Models for USA's $CO_2$ emission

| Model | MAE (Mt $CO_2$) | RMSE (Mt $CO_2$) | $R^2$ | MAPE (%) |
|---|---|---|---|---|
| ARIMA | 192.00 | 229.92 | 0.757 | 4.07 |
| SARIMA | 192.53 | 229.45 | 0.758 | 4.07 |
| Holt-Winters | 160.80 | 190.26 | 0.822 | 3.40 |

For the USA, Holt–Winters also gives the best fit, capturing the peak-and-decline structure with roughly 3-4% average error.

Table IV: Metric Evaluation of Time-Series Models for China's $CO_2$ emission

| Model | MAE (Mt $CO_2$) | RMSE (Mt $CO_2$) | $R^2$ | MAPE (%) |
|---|---|---|---|---|
| ARIMA | 518.86 | 600.49 | 0.924 | 4.84 |
| SARIMA | 515.80 | 599.19 | 0.924 | 4.81 |
| Holt-Winters | 510.00 | 585.50 | 0.926 | 4.72 |

All three models perform well for China, but Holt–Winters again yields the lowest MAE/RMSE and highest $R^2$.

Table V: Metric Evaluation of Time-Series Models for Brazil's $CO_2$ emission

| Model | MAE (Mt $CO_2$) | RMSE (Mt $CO_2$) | $R^2$ | MAPE (%) |
|---|---|---|---|---|
| ARIMA | 518.86 | 600.49 | 0.924 | 4.84 |
| SARIMA | 515.80 | 599.19 | 0.924 | 4.81 |
| Holt-Winters | 510.00 | 585.50 | 0.926 | 4.72 |

Given the shared numerical pattern, Holt–Winters is again selected as the most accurate model.

Table VI: Metric Evaluation of Time-Series Models for Russia's $CO_2$ emission

| Model | MAE (Mt $CO_2$) | RMSE (Mt $CO_2$) | $R^2$ | MAPE (%) |
|---|---|---|---|---|
| ARIMA | 51.07 | 59.17 | -2.195 | 3.07 |
| SARIMA | 50.98 | 59.11 | -2.188 | 3.07 |
| Holt-Winters | 53.78 | 61.37 | -2.437 | 3.23 |

For Russia, the near-constant emissions over the short test window lead to negative $R^2$ values even for models with low MAPE, a known artefact when variance around the mean is small.

### *D. Long-Term $CO_2$ Emission Projections (2024–2060)*

Using optimal models retrained on 2000–2023 data, projections to 2060 highlighted divergent pathways, with 95% prediction intervals accounting for uncertainty using the best performing models for each country is shown below:

Table VII: Metric Evaluation of the best-performing Time-Series Models for each's $CO_2$ emission

| Country | Best Performing Model | $R^2$ | MAE | RMSE | MAPE (%) |
|---|---|---|---|---|---|
| Nigeria | Holt-Winters | 0.803 | 8.31 | 9.84 | 6.35 |
| USA | Holt-Winters | 0.822 | 160.80 | 190.26 | 3.40 |
| China | Holt-Winters | 0.926 | 510.00 | 585.50 | 4.72 |
| Brazil | Holt-Winters | 0.926 | 510.00 | 585.50 | 4.72 |
| Russia | SARIMA | -2.188 | 50.98 | 59.11 | 3.07 |

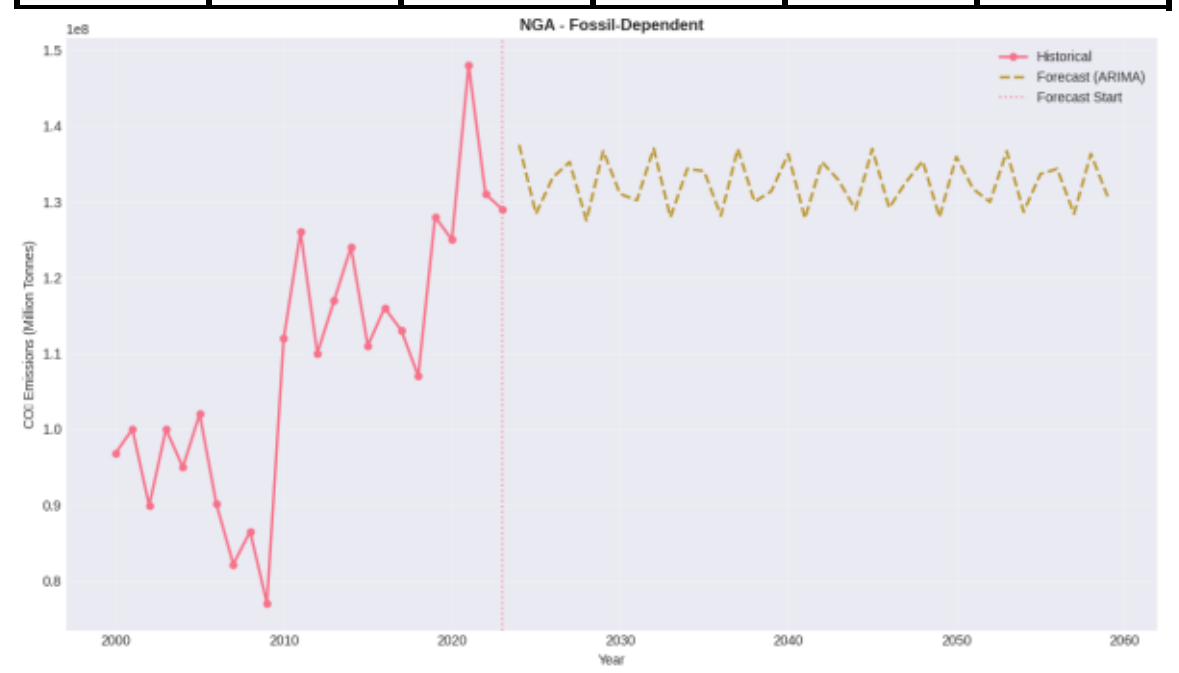

Figure 11: $CO_2$ forecast in Nigeria from 2024 to 2060

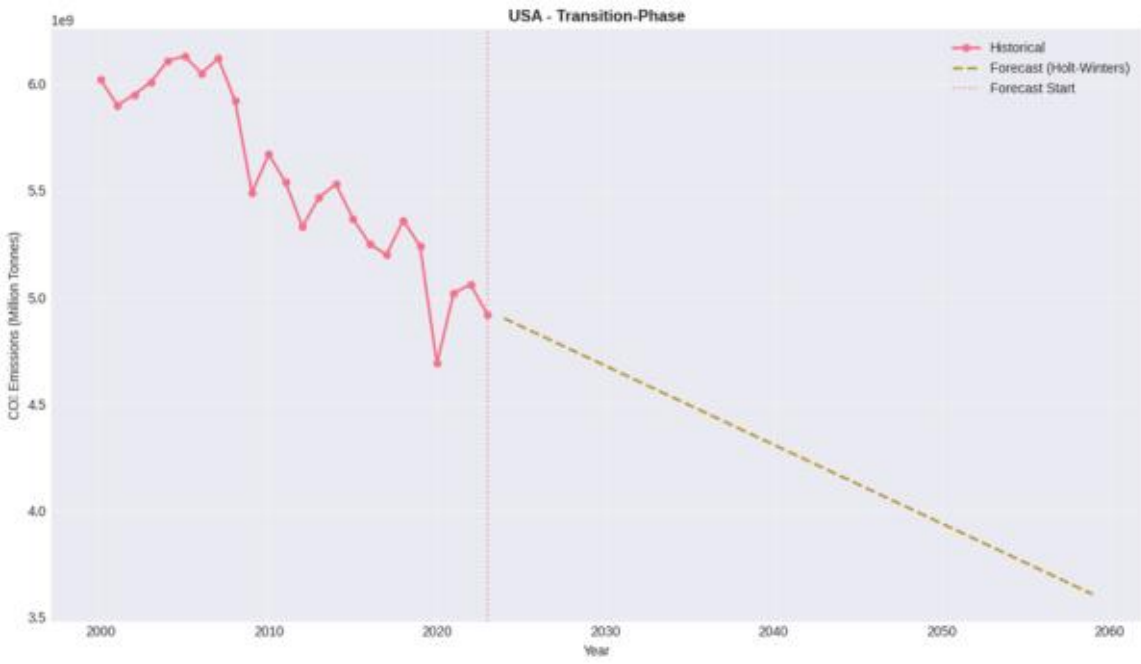

Figure 12: $CO_2$ forecast in USA from 2024 to 2060

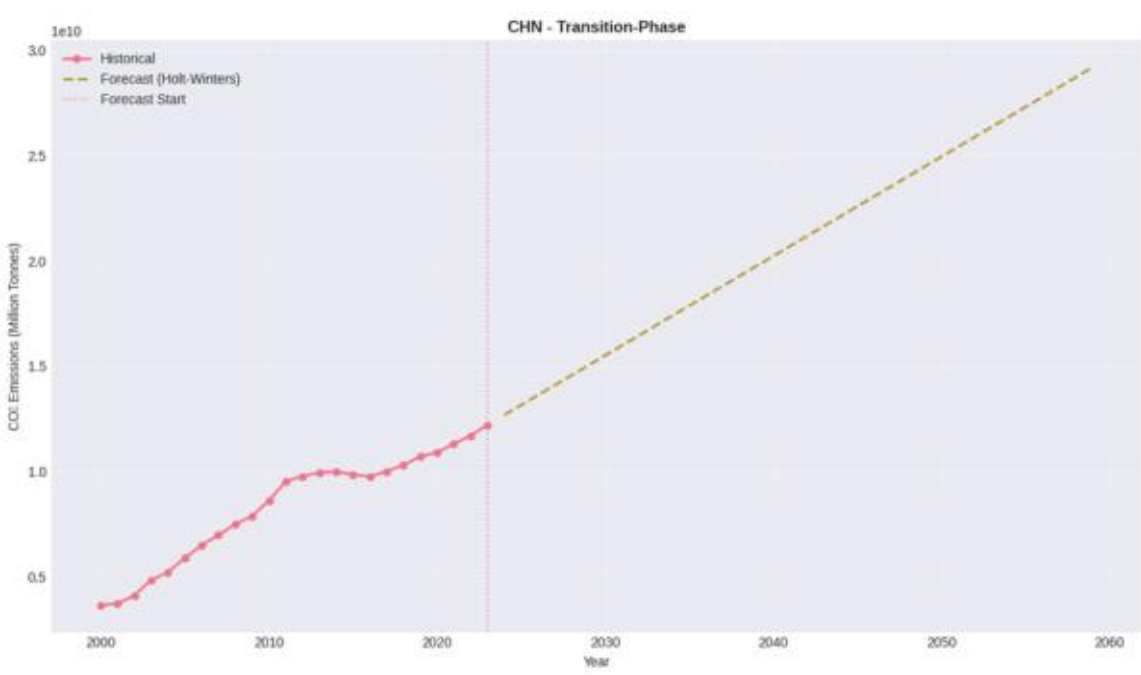

Figure 13: $CO_2$ forecast in China from 2024 to 2060

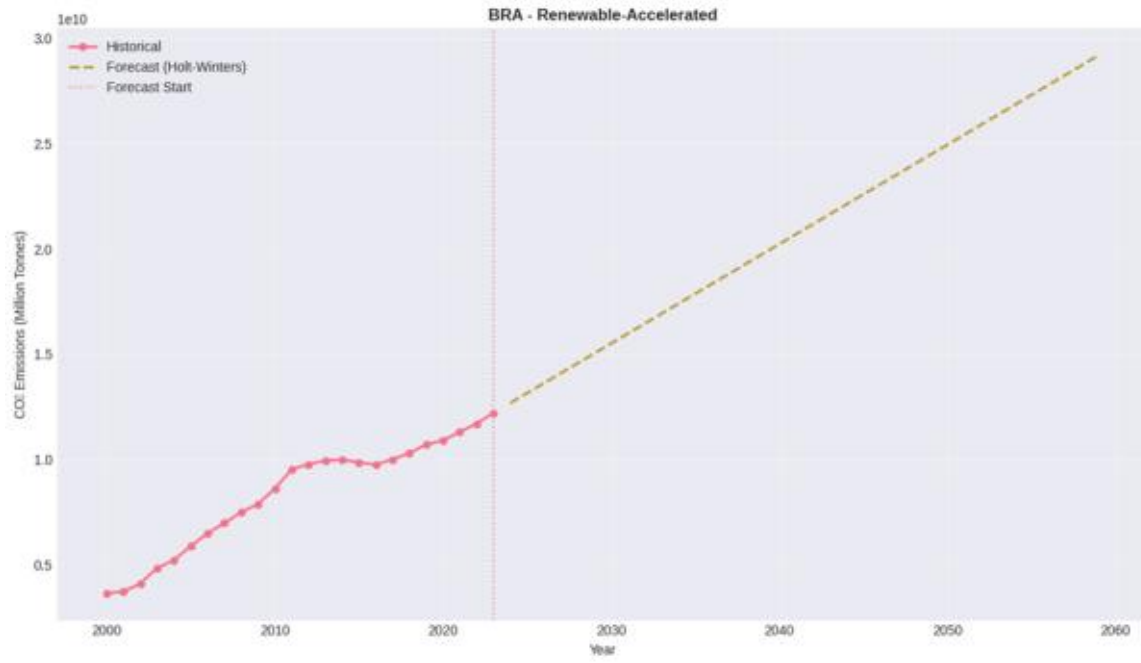

Figure 14: $CO_2$ forecast in Brazil from 2024 to 2060

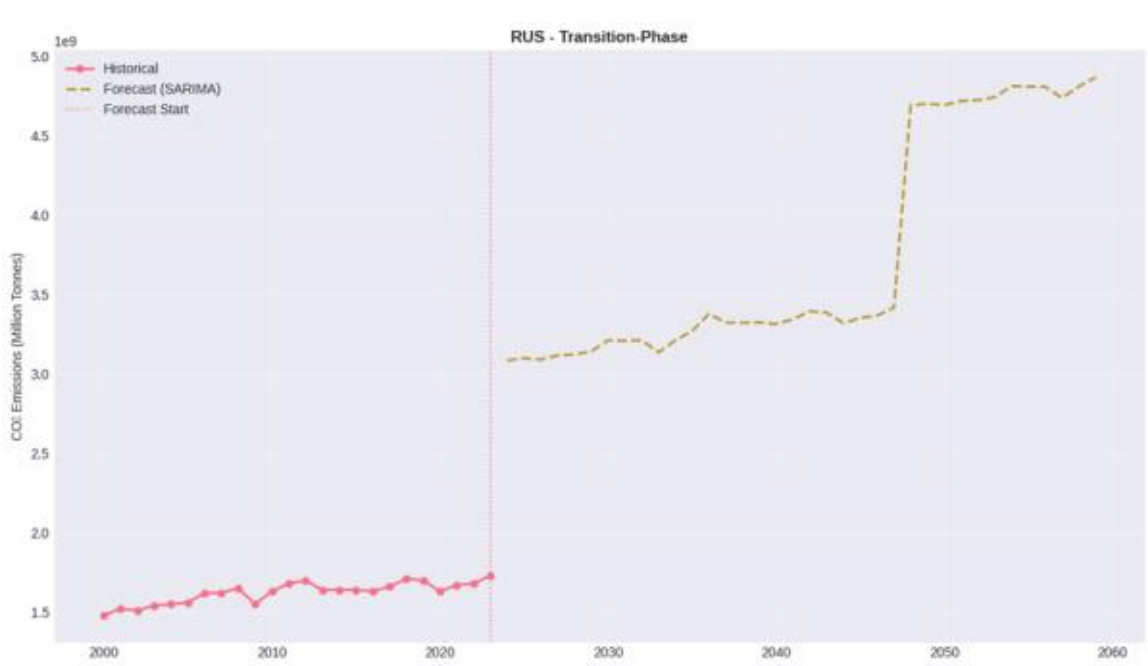

Figure 15: $CO_2$ forecast in Russia from 2024 to 2060

## VI. CONCLUSION

This study provides a comparative assessment of long-term $CO_2$ emission trends across five countries with distinct energy-system structures. By combining energy-mix profiling with time-series forecasting, the analysis demonstrates that national decarbonization prospects are strongly shaped by the balance between fossil fuels, renewables, and nuclear energy. Countries with fossil-dependent systems, such as Nigeria, show steadily increasing emissions that diverge from global net-zero pathways. Transition-phase economies, including the United States and China, exhibit declining or plateauing trajectories but still require substantial policy acceleration to achieve their long-term climate targets. Brazil's renewable-accelerated energy system supports a comparatively favorable emissions pathway, driven by extensive hydroelectricity and expanding bioenergy and wind resources. Russia remains largely constrained by its hydrocarbon-based export economy, resulting in flat emissions and limited progress toward decarbonization.

The forecasting results further reveal that Holt–Winters exponential smoothing effectively captures long-term trends for most countries, while SARIMA performs best in contexts with limited variability, such as Russia. These model-derived projections underscore the critical need for coordinated policy interventions, expanded renewable deployment, and structural energy reforms. Future research should explore scenario-based modelling, incorporate sector-specific emissions, and integrate machine learning models to better capture nonlinear transitions. Overall, the findings highlight that aligning national energy systems with net-zero objectives will require sustained and transformative efforts, particularly in regions where fossil fuel reliance remains deeply entrenched.

## REFERENCES

[1] IPCC, Climate Change 2022: Mitigation of Climate Change, Working Group III, Ch. 6, Cambridge Univ. Press, 2022.

[2] J. Jiang, S. Shi and A. Raftery, "Mitigation efforts to reduce carbon dioxide emissions and meet the Paris Agreement have been offset by economic growth," Commun. Earth Environ., vol. 6, no. 823, Oct. 2025.

[3] L. Maizland, C. Fong, "Paris to Kyoto: The History of UN Climate Agreements," Council on Foreign Relations, Nov. 2025

[4] International Energy Agency (IEA), World Energy Outlook 2023, International Energy Agency, Paris, 2023

[5] A. Cherp, J Jewell, V. Vinichenko, N. Bauer, E. De Cian, "Global energy security under different climate policies, GDP growth rates and fossil resource availabilities," Climate Change, vol. 1, pp. 136, 2013.

[6] V. Bektaş, "Structural Determinants of Greenhouse Gas Emissions Convergence in OECD Countries," Sustainability, vol. 17, no. 19, p. 8730, Sept. 2025.

[7] U.S. Energy Information Administration (EIA), "What is U.S. electricity generation by energy source?", EIA Today in Energy, Feb. 2024.

[8] P. Crompton and Y. Wu, "Energy consumption in China: past trends and future directions," Energy Economics, vol. 27, no. 1, pp. 195–208, Jan. 2005, doi: 10.1016/j.eneco.2004.10.006.

[9] Brandão, S. Q. et al. "Hydropower Enhancing the Future of Variable Renewable Energy Sources in Brazil." Energies, vol. 17, no. 13, 2024, Art. 3339.

[10] Mitrova, T. "Energy and the Economy in Russia." In The Russian Economy, edited volume, Springer, 2022.

[11] Edomah, N., Foulds, C., & Jones, A. "Energy Transitions in Nigeria: The Evolution of Energy Infrastructure Provision (1800–2015)." Energies, vol. 9, no. 7, 2016, Art. 484.

[12] I. Ayaz, "Forecasting carbon dioxide emissions using Machine Learning Methods: Turkey Example and Future Trends," DergiPark, 2024.

[13] M. Al-Lami and Á. Török, "Regional forecasting of driving forces of CO2 emissions of transportation in Central Europe: An ARIMA-based approach," Energy Reports, Volume 13, 2025, Pages 1215-1224, ISSN 2352-4847

[14] M. Mustafa, M. Akter, and S. Rahman, "$CO_2$ emissions forecasting using ARIMA, SARIMAX, ANN, RF, and LSTM: Evidence from Bangladesh," IEOM Society Proceedings, 2024.

[15] A.T. Mustafa, O.S.A. Al-Yazbaky, "Forecasting energy demand and generation using time series models: A comparative analysis of classical, grey, fuzzy, and

intelligent approaches," Franklin Open, Volume 12, 2025, 100350, ISSN 2773-1863

[16] Begum, A.M., Mobin, M.A. "A machine learning approach to carbon emissions prediction of the top eleven emitters by 2030 and their prospects for meeting Paris agreement targets." Sci Rep 15, 19469 (2025).